# MOTION BACKWARDS IN TIME?

# A FIRST EXAMPLE

Evangelos Chaliasos

365 Thebes Street

GR-12241 Aegaleo

Athens, Greece

It is of great significance the recent discovery (1998) by astronomers [1-4], found that the Universe not only could be open (contrary to the prevailing inflation theory, requiring a flat Universe), but perhaps it should have been expanding with a lower velocity in the past than today! In this regard, it is characteristic the statement of Brian Schmidt: "My own reaction is somewhere between amazement and horror"[2]. This "discovery" was in fact the interpretation given to the results of observations of type Ia supernovae at great distances from us. They found in fact that the supernovae were receding at a rate of expansion slower than its present value. And, since it was thought that their light was coming from the remote past, it was therefore "concluded" that the Universe, accelerating its expansion instead of decelerating it, is expanding today with a speed greater than that of the past. This "conclusion" is in a severe conflict to the Standard

Theory of the Big-Bang. Thus, in order to get out of this contradiction, physicists developed two possible alternatives.

According to the first one they gave the phenomenon the explanation that it is about an "antigravity" force which accelerates the expansion, contrary to the tendency of the material content of the Universe to decelerate it. They formally try to "explain" this "antigravity" force through the introduction of the cosmological constant $\lambda$ in the equations [5-7]. In fact, after the introduction of $\lambda$, we obtain the equation [8]

$$\Omega_0 = 2q_0 + (2/3)\lambda(c^2/H_0^2),$$

with the symbols having the usual meaning. Thus, "it is possible to have negative $q_0$, that is, an accelerating expansion, if $\lambda > 0$. This is because the $\lambda$-term introduces a force of cosmic repulsion". But, nevertheless, the Standard Model is too strongly founded, not only theoretically but observationally too, in order to simply reject it and substitute it with another model.

The second alternative is just to accept an open Universe and modify the prevailing inflation theory. For the latter see for example [9] for a brief account. For the modified one see [10-11], among others. In this second case it is assumed that our Universe is only one bubble-universe from many bubble-universes developed in a flat over-all-Universe in which the ordinary inflation took place. In everyone of the bubble-universes then a second inflation scenario took place, which leads $\Omega$ from 0 to 1 as a limiting case, so that the bubble-universes, and among these our own Universe, are open.

The whole problem emerged is so complicated, that many people expect it to be resolved only in the framework of a final theory (see for example [12]). It is really about a puzzling situation.

Thus the despair is diffuse among cosmologists. But let us hope. There is an exit from the above dilemma. The crucial point is the *arbitrary* acceptance of the fact that the light of the supernovae is coming from the past! Thus, the above problem *forces* us to accept exactly that their light is coming from the future! Then there is no dilemma at all between the Standard Model and the above discovery. Because, if really the light of the supernovae is coming from the future, it must then be expected to observe slower rate of expansion, which is what right the Standard Model predicts for the future.

In a theory of time reversal proposed by the author but not published yet, time is symmetric, resulting in a symmetry of the Universe too. Thus it is possible that the Universe is divided into two equal parts, the World and the Anti-World, consisting of matter and antimatter respectively. These, because of homogeneity of the whole Universe, must be "mixed" in the large scale, that is there must be islands of matter and (anti-) islands of antimatter uniformly distributed in the Universe, constituting the World and the Anti-World respectively. The basic characteristic, according to this theory, of the World is that time elapses normally (as we know it), but, instead, in the Anti-World time elapses in the opposite direction. Thus, finally, all physical phenomena evolve forward in time in the World, but they evolve backwards in time in the Anti-World.

In this way, the phenomenon we are dealing with, namely the reception of light signals from the future, is, at the same time, about the discovery of the Anti-World! Which means exactly the confirmation of this theory of time reversal! This is the case, because, since all physical phenomena are time-reversed concerning the Anti-World, it is then expected that light from the Anti-World must move backwards in time.

It is true that, according to this theory, light moving backwards in time must carry negative energy, and by means of this fact we would be perhaps able to distinguish it from light moving forward in time, which carries positive energy. But this is, unfortunately, not the case. Because what we are in fact measuring is energy flux $\Phi$, defined by $\Phi = dE/dt$, where $dE$ is the amount of energy absorbed by the detector during the time interval $dt$. Thus, since a negative $dE$ corresponds to an also negative $dt$, in contrast to a positive $dE$ corresponding to an also positive $dt$, there results no change in $\Phi$, that is there is no difference between those two cases.

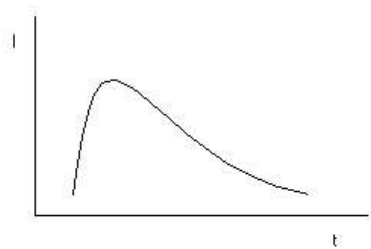 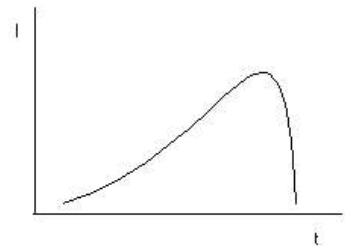

Fig. 1: light curve of a supernovaFig. 2: light curve of an anti-supernova

Nevertheless, we can claim that we can distinguish between the two cases. Thus, we can predict that exactly the light curve (fig. 2) of the above described "anti-supernova" must have the opposite time ordering from that (fig. 1) of the usual supernova, in other words they must be symmetric to each other. This must be the case because exactly of the time-reversal of all physical phenomena in the Anti-World. Of course it is a little difficult to construct a reliable light curve because exactly of the unexpected character of the phenomenon and the short duration of it.